\documentclass[preprint2]{aastex631}

\usepackage{graphicx}

\usepackage{orcidlink}

\usepackage{amsmath}

\newcommand{\imgsize}{501}

\newcommand{\numrandom}{100}
\newcommand{\psfwidth}{151}

\newcommand{\proftrunc}{5}
\newcommand{\corrnoisefwhm}{2}

\newcommand{\rawpsfurl}{https://irsa.ipac.caltech.edu/data/SPITZER/docs/irac/calibrationfiles/psfprf/Warm.IRAC.Extended.PSF.5X.091113.tar.gz}

\newcommand{\inputpixelscale}{0.60}

\submitjournal{ApJ}

\shorttitle{Stellar Populations of SPRC047}
\shortauthors{Laine et al.}

\begin{document}

\title{Stellar Population Properties in the Stellar Streams Around SPRC047}

\correspondingauthor{Seppo Laine}
\email{seppo@ipac.caltech.edu}

\author[0000-0003-1250-8314]{Seppo Laine} 
\affiliation{IPAC, Mail Code 314-6, Caltech, 1200 E. California Blvd., Pasadena, CA 91125, USA}

\author[0000-0003-3835-2231]{David Mart\'{i}nez--Delgado}
\affiliation{Instituto de Astrof\'{i}sica de Andalucía, CSIC, E-18080 Granada, Spain}

\author[0000-0002-8610-0672]{Kristi A. Webb}
\affiliation{Waterloo Centre for Astrophysics, University of Waterloo, Waterloo, ON N2L3G1, Canada}
\affiliation{Department of Physics and Astronomy, University of Waterloo, Waterloo, ON N2L 3G1, Canada}

\author[0000-0003-1710-6613]{Mohammad Akhlaghi}
\affiliation{Centro de Estudios de F\'isica del Cosmos de Arag\'on (CEFCA), Unidad Asociada al CSIC, Plaza San Juan 1, 44001, Teruel, Spain}

\author[0000-0001-5214-7408]{Roberto Baena--Gall\'{e}}
\affiliation{Universidad Internacional de la Rioja, Avenida de la Paz 137, E-26006 Logro\~{n}o, La Rioja, Spain}

\author[0000-0003-2922-6866]{Sanjaya Paudel}
\affiliation{Department of Astronomy, Yonsei University, Seoul 03722, Republic of Korea}
\affiliation{Center for Galaxy Evolution Research, Yonsei University, Seoul 03722, Republic of Korea}

\author[0000-0001-8428-7085]{Michael Stein}
\affiliation{Ruhr University Bochum, Faculty of Physics and Astronomy, Astronomical Institute (AIRUB), 44780 Bochum, Germany}

\author[0000-0002-8448-5505]{Denis Erkal}
\affiliation{Department of Physics, University of Surrey, Guildford GU2 7XH, UK}

\begin{abstract}

We have investigated the properties (e.g., age, metallicity) of the stellar populations of a ring-like tidal stellar stream (or streams) around the edge-on galaxy SPRC047 ($z = 0.031$) using spectral energy distribution (SED) fits to integrated broad-band aperture flux densities. We used visual images in six different bands and Spitzer/IRAC 3.6 $\mu$m data. We have attempted to derive best-fit stellar population parameters (metallicity, age) in three non-contiguous segments of the stream. Due to the very low surface brightness of the stream, we have performed a deconvolution with a Richardson--Lucy type algorithm of the low spatial resolution 3.6 $\mu$m IRAC image, thereby reducing the effect of the point-spread-function (PSF) aliased ``emission'' from the bright edge-on central galaxy at the locations of our three stream segments. Our SED fits that used several different star formation history priors, from an exponentially decaying star formation burst to continuous star formation, indicate that the age--metallicity--dust degeneracy is not resolved, most likely because of inadequate wavelength coverage and low signal-to-noise ratios of the low surface brightness features. We also discuss how future deep visual--near-infrared observations, combined with absolute flux calibration uncertainties at or below the 1 per cent level, complemented by equally well absolute flux calibrated observations in ultraviolet and mid-infrared bands, would improve the accuracy of broad-band SED fitting results for low surface brightness targets, such as stellar streams around nearby galaxies that are not resolved into stars.

\end{abstract}


\section{Introduction} \label{sec:intro}

The detection of stellar streams around nearby edge-on disk galaxies \citep[e.g.,][]{martinez2010,martinez2023b} provides direct, unequivocal evidence of the importance of minor mergers and dynamically significant satellite galaxy accretion events that will drive the growth and morphological evolution of massive disk galaxies until the current epoch. In the last decade, the main limitation that has prevented a better understanding of the actual contribution of these minor merger events in the build-up of stellar halos has been the missing accurate photometry of stellar streams. The detection of these faint structures is very challenging due to their low surface brightness (typically fainter than 26 AB mag~arcsec$^{-2}$). For this reason, the majority of known stellar streams have been discovered around nearby spiral galaxies using deep imaging and a very broad-band (``luminance'') filter \citep[e.g.,][]{martinez2015}. The use of a very broad-band filter does not allow the determination of stream colors, and thus, their stellar populations. However, the new generation of large-scale surveys (e.g., Dark Energy Spectroscopic Instrument or DESI Legacy Imaging Surveys; \citeauthor{Dey2019} \citeyear{Dey2019} and Ultraviolet Near-Infrared Optical Northern Survey or UNIONS\footnote{https://www.skysurvey.cc/}) has made it possible to obtain, for the first time, broad-band photometry of a few hundreds of these faint structures (e.g., \citeauthor{martinez2023b} \citeyear{martinez2023b}, Mir\'{o}--Carretero et al., in preparation). Interestingly, the colors of stellar streams around Milky Way analogs are redder than the survivor satellites, as determined by the Satellites Around Galactic Analogs (SAGA) survey \citep{geha2017,mao2021}, a difference that is still controversial. 

Photometric studies in the infrared have only covered a couple of nearby galaxies, including M83 \citep{barnes2014} and NGC 5907 \citep{laine2016}. The exact nature of these accreted, and in the process disrupted, companion galaxies can only be determined if some basic properties of the stellar populations of the disrupted companion galaxies, such as the age, metallicity, and mass of the dominant population of their constituent stars, are known. One complicating factor is the well-known stellar population age--metallicity degeneracy \citep[e.g.,][]{worthey1994}, exacerbated by the unknown amount of dust within the disrupted secondary galaxy. As demonstrated by, e.g., \citet{spitler2008}, \citet{laine2016}, and \citet{pandya2018}, visual--3.6~$\mu$m colors and spectral energy distribution (SED) fitting provide a new tool to make progress in resolving the age--metallicity degeneracy without obtaining expensive spectra. Consequently, a more accurate determination of the stellar mass-to-light $(M/L$) ratio and stellar mass will become possible. However, the power of SED fitting has not been tested thoroughly in the lowest surface brightness regime, including spatially unresolved faint tidal stellar streams around nearby disk galaxies.

In the Milky Way and Andromeda (M31) galaxies, stellar streams can be resolved into their individual stars, which has allowed for numerous studies of their stellar populations \citep[e.g.,][]{deboer2015,dey2023}. However, for streams around more distant galaxies this is not possible, and we can only study them with integrated surface brightness photometry. Without going to the detailed age and metallicity estimates, a few papers \citep[e.g.,][NGC~922]{martinez2023a} have used broad-band colors of the stellar tidal features to derive mass-to-luminosity ratios. \citet{laine2016} presented a study of sections of stellar streams around the nearby ($z$ = 0.002) edge-on galaxy NGC~5907. They used the Flexible Stellar Population Synthesis ({\sc fsps}) package \citep{conroy09,conroy10} to model the observations that covered wavelengths from the visual to 3.6 $\mu$m, and found evidence for a fairly massive disrupted companion galaxy, based on an old and metal-rich stellar population. \citet{foster2014} studied the narrow stellar stream apparently emanating from the Umbrella Galaxy NGC~4651 with Subaru Suprime-Cam \citep{miyazaki2002} wide-field camera observations in three broad bands ($g^{'}$, $r^{'}$, and $i^{'}$), and made estimates of the age and metallicity of the stream based on the visual color indices of the narrow stellar stream.

While traditionally only single-burst, exponentially declining star formation (SF) models have been used to fit stellar populations in low surface brightness targets, such as ultra-diffuse galaxies \citep[UDGs; e.g.,][]{pandya2018,buzzo2022}, nonparametric models could potentially provide a viable alternative for determining the star formation history (SFH) from SED fits. \citet{leja2019} recently demonstrated the advantages of nonparametric SFH models through ``ground-truth'' analysis of mock galaxies (see also \citeauthor{carnall2019a} \citeyear{carnall2019a} for a comparison of parametric SFH models, such as exponentially declining models). The nonparametric models offer more flexibility in describing realistic SFHs, and often produce less biased results with respect to parameters such as stellar mass and star formation rate (SFR). In comparison, parametric models are known to have biased trends between degenerate parameters (e.g., metallicity, dust, and age), given the inflexibility of their SFH prescription.

An unresolved problem (and not much discussed in any previously published work) is how much ``flexibility'' matters when the observations provide little constraint on the shape of the SFH (e.g., because the observations are of low signal-to-noise [S/N] or are limited in wavelength coverage). \citet{leja2019} concluded that SED fitting results based on low S/N and/or insufficient wavelength coverage are highly influenced by the model. Their recommendation was to use a model which is commensurate with the true SFH --- but of course the true SFH is not always known a priori! So the problem becomes circular. With this in mind, our approach has been to use an SFH model that is logically appropriate for the object we are studying. Therefore, e.g., a delayed-exponential SFH is not necessarily expected to be a bad choice for an object that perhaps stopped forming stars a long time ago. 

Any conclusions drawn from parametric models with respect to degenerate parameters should consider the expected biases discussed in \citet{leja2019}. A nonparametric continuous SF prior would actually be a poor prior for an SFH which quenched a long time ago (or even recently, as shown by \citeauthor{leja2019} \citeyear{leja2019}). The fact that some works use the continuous SF prior to fit all the galaxies, and that all the results are compared without consideration of the prior-bias, has been a problem in the SED fitting field in recent years.

\citet{lower2020} compared masses (and ages) derived from mock galaxy observations at $z = 0$, using S/N ratio $\approx$~30 photometry that covered far-ultraviolet (FUV) -- 0.5~mm wavelengths, fitted with several different SFH models. By comparing the true and recovered mass-weighted ages, they showed that while the derived ages from the nonparametric (with a Dirichlet prior with $\alpha$ = 0.7) model are typically overestimated, the parametric model derived ages are more typically underestimated. More generally speaking, a poor choice of an SFH model (one that is incommensurate with the true SFH) will lead to particularly biased results. For example, when assuming a constant SFH for a largely quiescent galaxy, or an exponentially declining SFH for a constantly star-forming galaxy. The specifics of the mass-weighted age bias were not studied in detail by \citet{lower2020}, as, unfortunately, they did not distinguish between the results for star-forming and quiescent mock galaxies. We suspect that the age-bias from the parametric models is less significant for quiescent galaxies. \citet{pacifici2023} recently examined the variance in SED fitting results using a variety of codes. Their paper provides a nice methodology for assessing SED fitting results, which involves critically comparing the posterior distributions against the model priors. We adopt this methodology.

We found an ideal target to test SED fitting in the ultra-low surface brightness regime in Canada–France–Hawaii Telescope (CFHT) images as a side product of a search for nearby interacting dwarf galaxies \citep{Paudel23} in the Sloan Digital Sky Survey (SDSS) and Legacy Surveys \citep{Dey2019}. This is a rare ring-like tidal stellar stream (originally cataloged and classified as a polar-ring galaxy by \citeauthor{moiseev11} \citeyear{moiseev11}) around the edge-on galaxy SPRC047 ($R.A.$ = 13$^{h}$59$^{m}$41$^{s}$, $Dec.$ = 25$^{\degr}$00$^{\arcmin}$45$^{\arcsec}$ at redshift $z$ = 0.03117; \citeauthor{huchra12} \citeyear{huchra12} and using $z~=~v/c$). The redshift of this galaxy is high enough so that there are not many foreground stars that would be projected on the stellar streams and that would alias emission from their extended point-spread-function (PSF) wings into parts of the stellar stream, but the galaxy is also close enough so that the stream can still be clearly defined (``spatially resolved''). These kinds of complex, multiple ring streams are the most suitable targets for N-body model fitting, with the ultimate goal of deriving the dark matter profiles of their host galaxies \citep{amorisco2015}. Also, observing systematic variations in the SED or color in the different parts of such stellar streams could be further employed in the future to study their formation. 

The rest of this article is organized as follows: In Section~\ref{sec:obs} we give details of the observations that we used in our study and explain how we derived uncertainties in the measured AB magnitudes. Section~\ref{sec:deconv} gives details on the deconvolution method (and its uncertainties) that we used to eliminate the aliasing of extended PSF emission from the central bright edge-on galaxy onto the faint stellar streams in the {\it Spitzer}/IRAC 3.6 $\mu$m image. Section~\ref{sec:SED} summarizes how we performed the SED fits with our chosen SED-fitting code, {\sc prospector}. Next we show how we derived our results (Section~\ref{sec:results}), discuss the results and their implications in Section~\ref{sec:discussion}, and finally summarize our main findings in Section~\ref{sec:conclusions}. Throughout this work, we adopt a flat Lambda cold dark matter ($\Lambda$CDM) cosmology with $\Omega_{\rm m}$ = 0.27 and $H_{\rm 0}$ = 72 km~s$^{-1}$~Mpc$^{-1}$, and follow the AB magnitude system \citep{oke83}.

\section{Observations and Data Reduction} \label{sec:obs}

\begin{figure*}
\begin{center}
\includegraphics[scale=0.2]{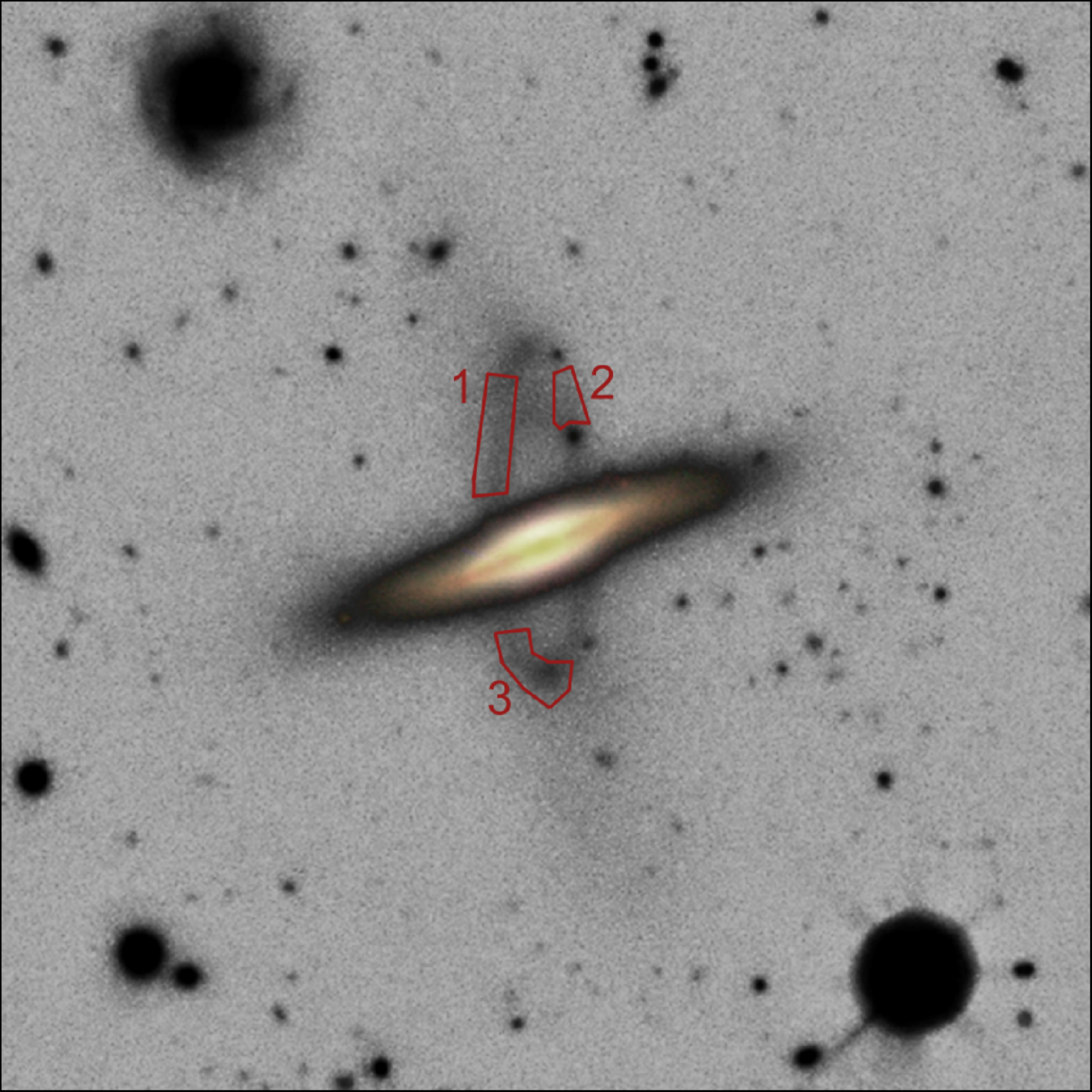}
\caption{Combined Legacy Survey \citep{Dey2019} $grz$-band tri-color image of SPRC047. The field of view shown is 100\arcsec$\times$100\arcsec or about 60 kpc$\times$60 kpc. The locations of the apertures where photometry was taken are also shown and numbered. Credit: Giuseppe Donatiello.\label{figopt}}
\label{optimage}
\end{center}
\end{figure*}

\subsection{IRAC 3.6 Micron Observations} \label{sec:IRAC_obs}

Spitzer/IRAC \citep{fazio2004, werner2004,https://doi.org/10.26131/irsa543} 3.6 $\mu$m (channel 1) near-infrared (NIR) observations of SPRC047 were taken on 2018 October 4 starting at 23:41 UT. Thirty medium-scale (median dither separation 53 pixels) cycling dither pattern 100 second frames were taken. The total exposure time was 30$\times$93.6~s = 2808~s.

We removed three frames from further analysis, as the target in these frames was covered by scattered light coming from a nearby star outside the field of view, leaving us with 27 100 s frames. Next we masked the long-term residual images in each frame created by preceding Spitzer/IRAC observations of bright targets and by the SPRC047 observation itself. A contributed Spitzer/IRAC software package {\sc imclean} \citep{jhora99_2021_4850526} was used to perform this step. 

Next we used the Spitzer custom mosaicking software {\sc mopex} with most of its default parameters, except that we set the top and bottom Rmask levels to 5 for outlier detection, used RM\_Thres = 0.5, and built a 3.6 $\mu$m mosaic of SPRC047. We then measured the ``sky background'' in five empty areas of sky close to SPRC047 and determined an average background level of $-0.0051$ MJy sr$^{-1}$, which was subsequently subtracted from all the pixels in the 3.6 $\mu$m image. Uncertainties for the IRAC 3.6 $\mu$m photometry are dominated by the deconvolution uncertainty of 0.05 AB mag (see Section~\ref{deconv:errors} for more information). No foreground Galactic extinction corrections were made to the measured IRAC 3.6 $\mu$m integrated flux densities, as such corrections are smaller than 0.01 AB mag. The final measured AB magnitude, and its uncertainty, are given in Table~\ref{table1}.

\subsection{Visible Light Observations} \label{Visual_obs}\label{sec:vis_obs}

We used images that contained SPRC047 from the CFHT archive\footnote{https://www.cadc-ccda.hia-iha.nrc-cnrc.gc.ca/en/cfht/.} and the Legacy Imaging Surveys \citep{Dey2019}. Visual observations in the $g$-, $r$-, and $i$-bands were found in the CFHT archive and in the $g$-, $r$-, and $z$-bands in the Legacy Imaging Surveys (see Figure~\ref{optimage}). We obtained pre-processed CFHT images that had undergone bias subtraction, flat-fielding, and flux calibration by the MegaCam pipeline system, {\sc elixir}. To further remove the sky background, we followed a method described in \citet{Paudel23}, in which we generated a background map by masking out identified sources in the image using Source Extractor \citep{Bertin1996} maps. The segmentation images were then filled with median values from surrounding pixels to eliminate light contributions from stars and background galaxies. The resulting background map was subtracted from the original FITS file to be used in aperture photometry.

The CFHT images have exposure times of about 420 seconds in the $g$ and $r$ bands and 280 seconds in the $i$ band. The Legacy Imaging Survey images have depths of about 435, 268, and 400 seconds in the $g$, $r$, and $z$ bands. Compared to, e.g., the Rubin's Legacy Survey of Space and Time that is expected to take 825 30-second exposures with a telescope twice the diameter of CFHT or Dark Energy Camera (DECam; it was used for our Legacy Survey images), the former will reach a statistical noise level that is about 30 times lower than in our images.

To obtain uncertainty estimates in visual photometry, we performed aperture photometry in five empty regions of the sky with the same aperture size as what was used for the photometry of the regions in the ring-like tidal stellar streams. We took the dispersion in the measurements of the empty regions to represent the uncertainty in photometry alone. The magnitude of these uncertainties was typically less than 0.05 mag in each band and aperture.

We also added an uncertainty due to sky background subtraction. We used the extreme measured sky background values in the five ``empty'' areas of the sky to subtract the sky and then performed photometry to obtain the aperture integrated fluxes. We then converted the flux difference from the extreme high and low background subtraction cases and converted it to AB magnitudes, to obtain a magnitude uncertainty due to background subtraction. The magnitude of the background subtraction uncertainty was typically 0.05 mag or less.

We also included an estimate of the uncertainty in the photometric calibration in the various bands. This was estimated to be 2 per cent in all the bands. The final total uncertainties are given in Table~\ref{table1}. All the measured magnitudes in visual bands were corrected for foreground Galactic extinction using \citet{schlafly2011} and Bayerstar\footnote{http://argonaut.skymaps.info/}.

\begin{deluxetable*}{lcccccc}[t]
\tablewidth{33pc}
\tablecaption{Measured (and adopted) AB magnitudes and their uncertainties. \label{table1}}
\tablehead{
\colhead{$g$ (CFHT)} & \colhead{$r$} (CFHT) &
\colhead{$i$} (CFHT)  & \colhead{$g$ (Leg. Survey)} &
\colhead{$r$ (Leg. Survey)} & \colhead{$z$ (Leg. Survey)} &
\colhead{IRAC 3.6 $\mu$m}
}
\startdata
\multicolumn{7}{c}{Region 1 ($\approx$~53 arcsec$^{2}$ or 19 kpc$^{2}$)} \\
20.68$\pm$0.04 & 20.05$\pm$0.08 & 19.71$\pm$0.03 & 20.67$\pm$0.05 & 19.94$\pm$0.03 & 19.48$\pm$0.06 & 20.08$\pm$0.05\\
\multicolumn{7}{c}{Region 2 ($\approx$~22 arcsec$^{2}$ or 8 kpc$^{2}$)} \\
21.69$\pm$0.07 & 21.10$\pm$0.12 & 20.74$\pm$0.04 & 21.69$\pm$0.06 & 21.00$\pm$0.06 & 20.50$\pm$0.05 & 21.33$\pm$0.05\\
\multicolumn{7}{c}{Region 3 ($\approx$~48 arcsec$^{2}$ or 17 kpc$^{2}$)} \\
20.60$\pm$0.04 & 19.92$\pm$0.07 & 19.87$\pm$0.03 & 20.57$\pm$0.05 & 19.87$\pm$0.04 & 19.42$\pm$0.05 & 19.96$\pm$0.05\\
\enddata
\end{deluxetable*}

\section{Deconvolution of the IRAC 3.6 Micron Image}\label{sec:deconv}

\subsection{Description of the Overall Procedure}\label{subsec:deconv:gen}

The distortions introduced into images by the acquisition process in astronomical observations are well known. One of the most challeging is the loss of spatial resolution introduced by the PSF. In the case of ground-based observations, the PSF is dominated by atmospheric turbulence, while for space telescopes the PSF is mostly set by the optics of the instrument. The classic equation that describes the image formation process is

\begin{equation}\label{imeq}
    \mathbf{i} = [\mathbf{o} * \mathbf{h}] \diamond \mathbf{n}.
\end{equation}

\noindent Here $\mathbf{o}$ is the object or unknown to be estimated, $\mathbf{i}$ is the image or data set acquired by the instrument, $\mathbf{h}$ is the PSF, and $\mathbf{n}$ is the noise. The symbol $*$ denotes a convolution operation that is a valid approximation when the PSF is assumed shift-invariant, and the symbol $\diamond$ is a pixel-by-pixel operation that reduces to addition when noise is additive and independent of $\mathbf{o} * \mathbf{h}$, while for shot noise it is an operation that returns a random deviate drawn from a Poisson distribution with a mean equal to $\mathbf{o} * \mathbf{h}$. Direct inversion of Equation (\ref{imeq}) in the Fourier domain cannot be made because of noise amplification in the vicinity of the spatial cutoff frequency. 

Equation (\ref{imeq}) is an ill-posed statement with non-unique stable solutions, and requires deconvolution approaches that generally involve non-linear optimization algorithms, to recover the original signal $\mathbf{o}$ from the PSF-convolved and noise-degraded data $\mathbf{i}$. If the PSF is well characterized, then it can be considered to be a deterministic variable, and static-PSF algorithms can be utilized that do not require an update of the PSF during the optimization process. 

Bayesian methodology is often used to find a solution compatible with the statistical nature of the signal that is sought. In this work, we have used the adaptive wavelet maximum likelihood estimator \citep[AWMLE;][]{Otazu:2001}. This is an iterative Richardson--Lucy-type algorithm that maximizes the likelihood between the image and the projection of a possible solution onto the data domain, considering a combination of the Poisson noise, intrinsic to the discrete nature of the signal, and the Gaussian readout noise of the electronic detector. AWMLE has been applied to survey-type data to improve the limiting magnitude and resolution by \cite{Fors:2006}, and for the reconstruction of adaptive optics (AO) observations of extended objects, in contrast to blind deconvolution algorithms \citep{baena:2013} that are \textit{a priori} more suited to dealing with the variability of AO-PSFs than static-PSF approaches.

AWMLE is fully described by \cite{baena:2011}, and has been tested for AO photometry problems, showing that its accuracy is comparable, or even better than those of other well-known approaches, such as the PSF-fitting StarFinder, especially when the PSF that is used to deconvolve the image does not depart from the PSF that creates the image more than 5 per cent in terms of the Strehl Ratio. The mathematical expression behind AWMLE is

\begin{equation}\label{awmle}
    \hat{o}^{(n+1)} = \mathrm{K} \hat{o}^{(n)} \Bigg[ h^{-} * \frac{\sum_j \big( \omega_j^{o_h^{(n)}} + M_j \big( \omega_j^{i{\acute{}}} - \omega_j^{o_h^{(n)}} \big) \big)}{h^{+} * \hat{o}^{(n)}}  \Bigg].
\end{equation}

\noindent where $\hat{o}^{(n)}$ is the estimated signal at iteration $n$, $h^{+}$ and $h^{-}$ are the PSFs that project the information from the object domain to the image domain and vice versa, respectively. The so-called direct projection $o_h$ is the object projected onto the data domain by means of the PSF, that is, $o_h = h^{+} * \hat{o}$. The variable $i{\acute{}}$ is a modified version of the image, and appears because of the explicit inclusion of the readout Gaussian noise into the Richardson--Lucy scheme \citep{Nunez:1993}, and represents a pixel-by-pixel filtering operation in which the original data set is replaced by this modified version. It must be calculated at each iteration $n$ since it depends on the projected object $o_h$ \citep{baena:2013}. The parameter K is a constant to preserve the energy to guarantee the relation $\Sigma_{xy}$$i(x,y)$ = $\Sigma_{xy}$$o$(x,y), and it is computed after each iteration. 

AWMLE adds the novelty of decomposing the data set $i{\acute{}}$\ by means of the wavelet transform and a pixel-based probabilistic mask $M_j$ to adapt the level of reconstruction of each wavelet plane to the automatically inferred presence of noise, i.e., the effective number of iterations does not have to be the same at each wavelet plane. This characteristic leads to some advantages in the treatment of information; for instance, high frequency features, such as the noise of point-like sources, may need a different level of reconstruction than wide and extended structures. Although this feature seems to be convenient in our case, we found in practice that reconstruction artifacts grew up iteration after iteration, and hence we opted to cancel the wavelet decomposition by setting $M_j$ to $1$ at all wavelet planes. In that case, Equation (\ref{awmle}) reduces to

\begin{equation}\label{awmle2}
    \hat{o}^{(n+1)} = \mathrm{K} \hat{o}^{(n)} \Bigg[ h^{-} * \frac{\sum_j  \omega_j^{i{\acute{}}}  }{h^{+} * \hat{o}^{(n)}}  \Bigg].
\end{equation}

\begin{figure*}
  \begin{center}
    \includegraphics[width=0.35\linewidth]{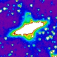}
    \includegraphics[width=0.35\linewidth]{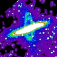}
  \end{center}
  \caption{\label{fig-deconv}\small IRAC 3.6 $\mu$m input (left) and deconvolved (right) images. North is up and east is left. The width and height are 151 $\times$ 151 pixels or about 90\arcsec $\times$ 90\arcsec.}
\end{figure*}

Therefore, the estimated object is directly compared with the modified data, since $i{\acute{}} = \sum_j  \omega_j^{i{\acute{}}}$. The standard deviation of the readout noise was estimated from source-free regions detected by the NoiseChisel software \citep{gnuastro, noisechisel19}. We used the published warm IRAC extended PSF\footnote{\url{\rawpsfurl}} that was rotated and scaled to align with the correct orientation of the image and at the same pixel scale ($\inputpixelscale$ arcsec~pixel$^{-1}$). The absence of reconstruction artifacts around the brightest star in the field of view showed that the deconvolution process was performed efficiently. The input and output images to and from the deconvolution can be seen in Figure~\ref{fig-deconv}.

As the IRAC 3.6 $\mu$m channel images have a PSF with a substantially larger full width half maximum (FWHM) than in the visual, and because this broad PSF effectively spreads emission to large distances, we decided to perform the deconvolution described above only on the 3.6 $\mu$m image to obtain a more accurate flux measurement, unaffected by the central edge-on galaxy. We discuss this point further below in Section~\ref{sec:photometry}.

\subsection{Measurement Uncertainties in the Deconvolved Image}\label{deconv:errors}

With the much higher spatial resolution 3.6 $\mu$m image obtained after the deconvolution, we can perform aperture photometry on any desired part of the image. The deconvolved image $\mathbf{o}$ is a proposed solution for the object that created the image $\mathbf{i}$, and is compatible with the degradation introduced by the instrument PSF and noise. Furthermore, that solution is always intrinsically positive, i.e., no negative values can exist \citep{starck2002}. However, model imprecisions, noise amplification, and artifacts created during the convergence process can make it difficult to define an uncertainty for a photometric measurement, in contrast to typical procedures applied to the image data $\mathbf{i}$.  Here we review the strategy we used to derive an uncertainty estimate for the aperture photometry from the deconvolved image\footnote{The scripts used to measure the flux density in the deconvolved image and to estimate its uncertainty are available at https://gitlab.com/makhlaghi/sprc047-stream-clumps-photometry.}.

\begin{figure*}
  \begin{center}
    \includegraphics[width=0.18\linewidth]{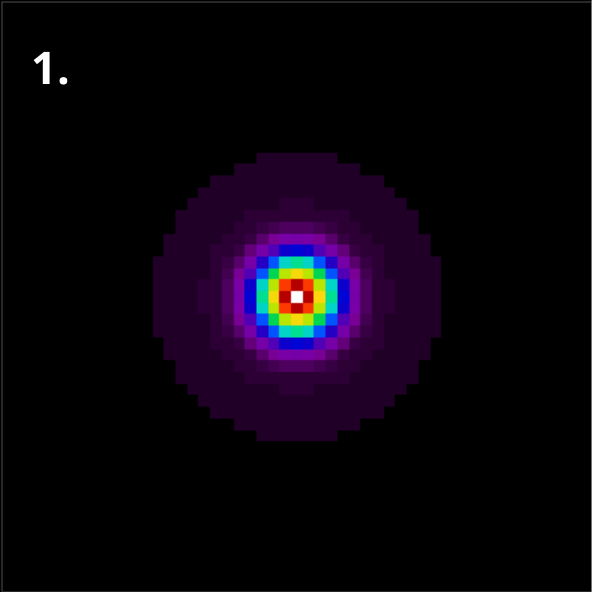}
    \includegraphics[width=0.18\linewidth]{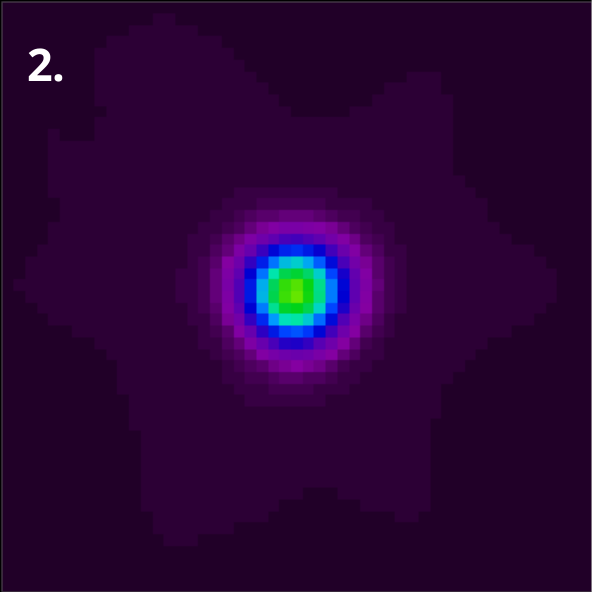}
    \includegraphics[width=0.18\linewidth]{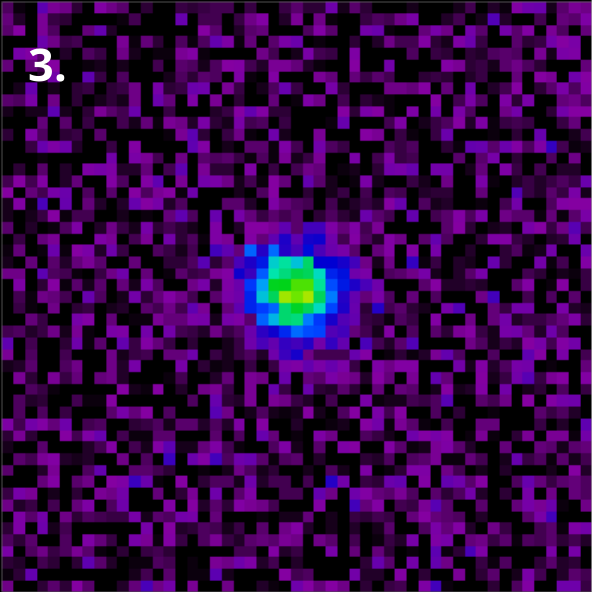}
    \includegraphics[width=0.18\linewidth]{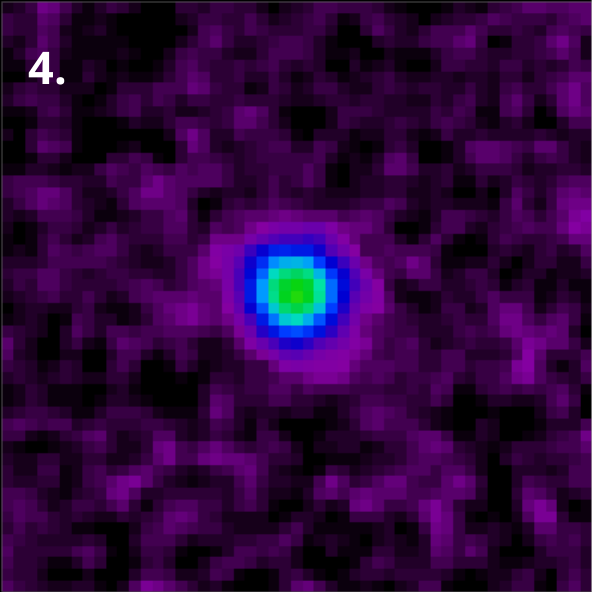}
    \includegraphics[width=0.18\linewidth]{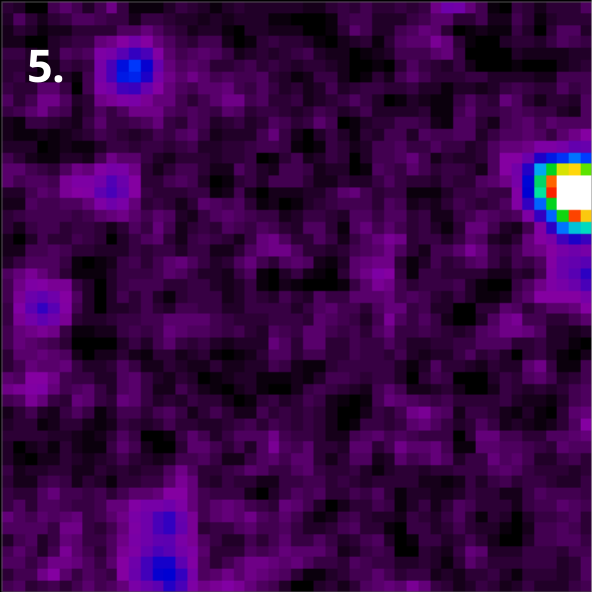}
  \end{center}
  \caption{\label{fig-err-intro}\small Mock noise estimation. From left to right: 1. the profile before convolution; 2. the same after convolution; 3. the same as 2. after adding raw noise; 4. the same as 3. after convolving the raw noise image with a sharp kernel to simulate correlated noise; 5. a crop from a relatively empty region of the image with the same crop size as image 4. shows the similarity of the real and mock correlated noise. All images have the same color scale. The width and height are 51 $\times$ 51 pixels or about 30\arcsec $\times$ 30\arcsec.}
\end{figure*}

First, we made a mock Gaussian profile with a FWHM of five pixels, truncated at $\proftrunc\times$FWHM. It is shown in Figure \ref{fig-err-intro} (panel 1). We then convolved the image with the published IRAC PSF described above. Also, because of the small size of our mock test image ($\imgsize\times\imgsize$), only the central $\psfwidth\times\psfwidth$ pixel region was used, since using the full PSF would have been too computationally expensive, with no added statistical precision; see Figure \ref{fig-err-intro} (panel 2). Furthermore, the mock image is mainly dominated by noise, hence making it difficult to obtain a stable estimate of the parameter K to conserve the energy during the convergence process. To help the algorithm to converge properly to a stable solution (set K to a ``fixed'' value after every iteration), two bright stars were placed at very distant positions from this test profile (taking care that they did not contribute at all to the pixel under study here). 

Setting the profile's ``total'' input magnitude would not give an accurate reference, because it is hard to measure in observed data. Therefore, after convolving the profile with the PSF, we scaled the convolved image (multiplied by a constant) such that within a $3\times3$ pixel box around the center, it has the same pixel sum as the main target in the observed image.

To simulate correlated noise, we first created ideal noise with a fixed $\sigma$ (where each pixel's noise is not affected by its nearby pixels) as shown in Figure \ref{fig-err-intro} (panel 3). Afterwards, we convolved the noise image with a sharp Gaussian kernel (FWHM$=$\corrnoisefwhm; Figure \ref{fig-err-intro}, panel 4). The $\sigma$ value of the ideal noise was determined by the following procedure: because we knew the noise of the final observed image, the ideal noise was manually adjusted such that after convolving the image with the sharp Gaussian kernel, the noise was similar to the observed level.

\begin{figure*}
  \begin{center}
    \includegraphics[width=0.18\linewidth]{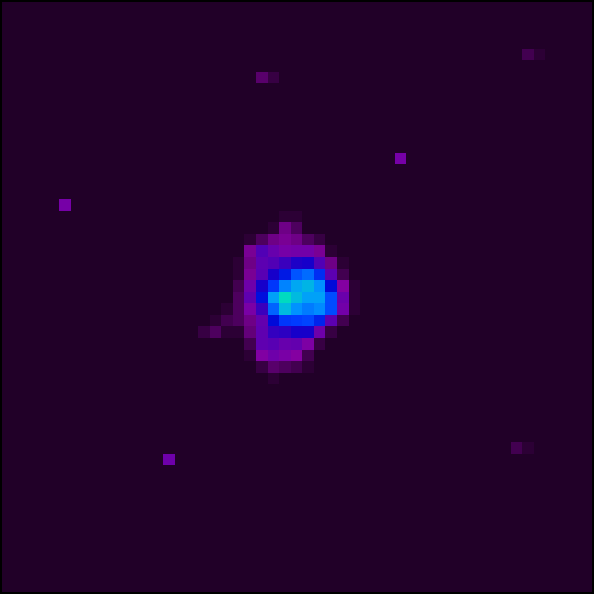}
    \includegraphics[width=0.18\linewidth]{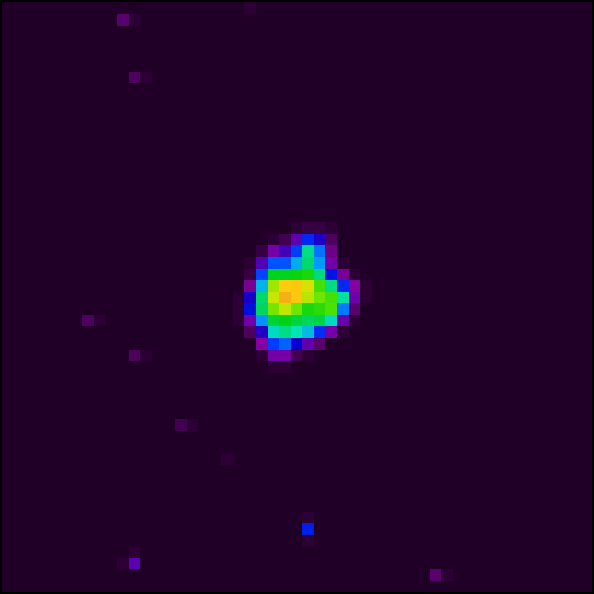}
    \includegraphics[width=0.18\linewidth]{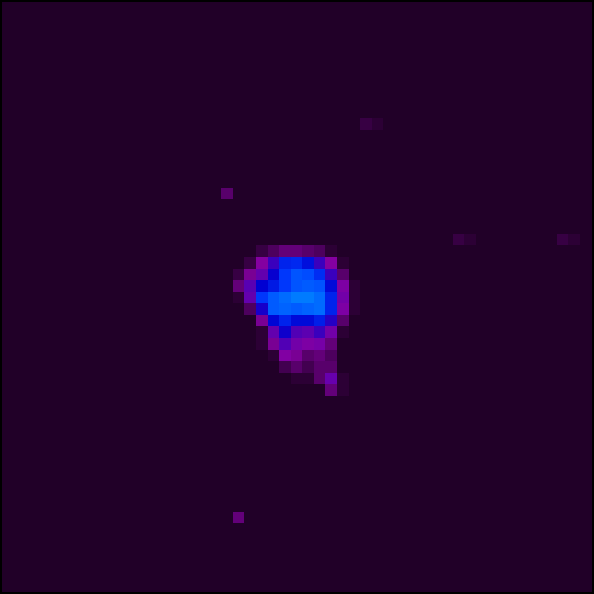}
    \includegraphics[width=0.18\linewidth]{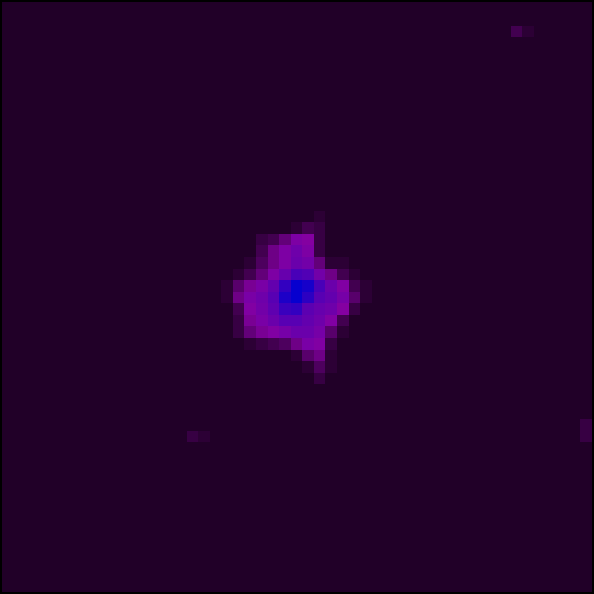}
    \includegraphics[width=0.18\linewidth]{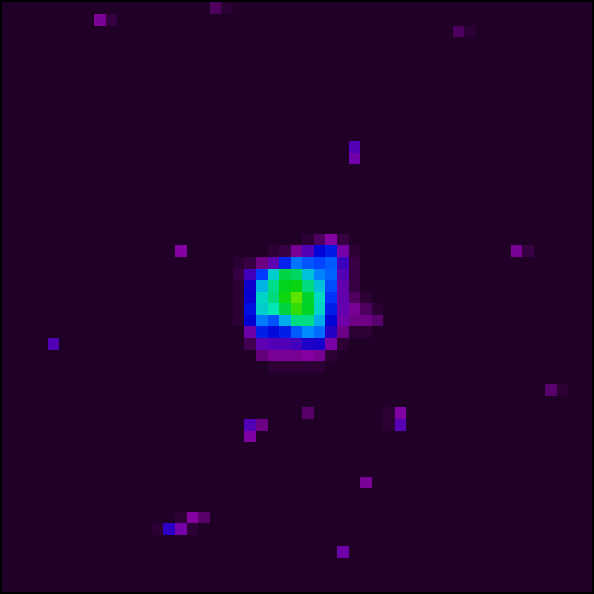}
  \end{center}
  \caption{\label{fig-err-demo}\small Five example deconvolved images. Color scale is the same as in Figure \ref{fig-err-intro}. The width and height are 51 $\times$ 51 pixels or about 30\arcsec $\times$ 30\arcsec.}
\end{figure*}

Having found the ideal noise $\sigma$, we simulated correlated noise on the convolved image $\numrandom$ times. Each time, we only changed the random number generator seed from 1 to $\numrandom$ (so that the exact same noise pattern could be reproduced). These 100 images were then fed into the deconvolution algorithm. The results for the first five images can be seen in Figure \ref{fig-err-demo}.

To measure the accuracy of the aperture-integrated flux density measurements in the deconvolved image, we measured the brightness within an aperture of five pixel radius. First, this was measured on the image without any noise and \emph{before convolution}; (Figure \ref{fig-err-intro}, panel 1). Then it was measured in the $\numrandom$ deconvolved images. From this basic experiment we concluded that we were able to measure the signal and its S/N. Because this test was done on a sharp-edged object, we expect the obtained uncertainty estimate to be an upper limit for diffuse source photometry, such as those measured in the three regions of the stellar stream of SPRC047 (see Section~\ref{sec:photometry}).

\section{SED Fitting} \label{sec:SED}

The modeled stellar populations in this study are based on the stellar population synthesis (SPS) models as implemented in the {\sc fsps} library \citep{conroy09,conroy10} with Modules for Experiments in Stellar Astrophysics (MESA) Isochrones and Stellar Tracks (MIST; \citealt{choi2016}; \citealt{dotter2016}, based on the {\sc mesa} stellar evolution code; \citealt{paxton2011, paxton2013, paxton2015, paxton2018}), and Medium resolution Isaac Newton Library of Empirical Spectra (MILES\footnote{http://miles.iac.es/.}) spectral templates \citep{sanchez2006}. A \citet{kroupa2001} initial mass function was assumed. The fully Bayesian inference code {\sc prospector}\footnote{https://github.com/bd-j/prospector.} \citep[v.1.0.0;][]{leja2017,johnson19a,johnson21} was used to generate SEDs from {\sc fsps} and compare the models to the observations. Because models are generated on the fly, the {\sc prospector} code allows for flexible model specification with larger numbers of parameters than are computationally tractable in typical grid-based searches.

One limitation of the {\sc fsps} models is that they do not include a metal enrichment history (we assumed a fixed stellar metallicity, with a prior on log[$Z$/$Z_{\odot}$] that was uniform with a minimum of $-2.0$ and maximum of $+0.2$ in all the fits) or non-solar abundances. The various metallicity evolution assumptions have been discussed by, e.g., \citet{bellstedt2020}. The effects of $\alpha$-element enhancement on broadband SEDs are expected to be much smaller than other effects. Another downside with {\sc prospector}/{\sc fsps} SED modeling, as with any Bayesian-based routine, is that the {\sc prospector} results (posteriors) are dependent on the product of the prior and the likelihood. Where the data are not sufficiently informative (i.e., do not constrain the parameter space of the available models), the prior will dominate the results. Thus, it is always important to feed the code with informative data and to run models with different prior assumptions to test and understand the dependencies so that plausible interpretations can be made.

The lack of observational constraints at ultraviolet (UV) or mid- to far-infrared wavelengths limits our ability to constrain the dust properties of the stellar stream around SPRC047. Moreover, the shape of the visual--NIR SED is degenerate between metallicity and dust extinction \citep[e.g.,][]{buzzo2022}. Given that several studies have demonstrated the sensitivity of SED-fitting results to the specification of the dust model and SFH models \citep[e.g.,][]{carnall2019a, leja2019, lower2020, salim2020}, a variety of models were explored in this work.

We adopt the two-component model for dust attenuation following \citet{charlot2000}, which separates the dust associated with young stars, and a second dust component uniformly screening all stars. The first component (dust around young stars) was not used in this study.

\noindent The uniform dust screen mimics diffuse dust, and has the following variable attenuation curve \citep{noll2009}:
\begin{equation}\label{eqn:dust_diffuse_2}
\begin{aligned}
 \tau_\mathrm{dust,~diffuse}(\lambda) =& 
 \frac{\hat{\tau}_\mathrm{dust,~diffuse}}{4.05} 
 \left( \frac{ \lambda }{ \text{5500~\AA} }\right)^{n}
 \\
 &\times
 \left( k^\prime(\lambda) +  D(\lambda) \right) 
\end{aligned}
\end{equation} 
\noindent where $n$ is the diffuse dust attenuation index, $k^\prime(\lambda)$ is the slope of the diffuse dust attenuation curve, and $D(\lambda)$ describes the UV bump. The free parameters of the dust attenuation model are the diffuse dust normalization constant, $\hat{\tau}_\mathrm{dust,~diffuse}$, and the dust attenuation index, $n$. Note that the visual extinction, $A_{V}$, equals 1.086~$\times$ $\tau_\mathrm{dust,~diffuse}$(550 nm). Two dust attenuation models were explored: 1) a \citet{calzetti2000} diffuse dust screen model without contribution from a UV bump, and 2) a \citet{kriek2013} model that links the slope of the diffuse dust attenuation curve, as set by $k^\prime(\lambda)$, to the strength of the UV bump. We assume $k^\prime(\lambda)$ = $-0.7$, the default in {\sc fsps}, in all our \citet{kriek2013} diffuse dust attenuation models. The Milky Way and the Small Magellanic Cloud attenuation curves are generally well covered by these options, although the \citet{calzetti2000} dust attenuation is above the normal galaxy dust attenuation levels at the wavelengths that we are interested in.

Dust emission was also included, in which the energy balance requirement propagates the energy attenuated by the dust to IR wavelengths. {\sc prospector} uses the templates from \citet{draine2007} that include a silicate--graphite polycyclic aromatic hydrocarbon (PAH) model of interstellar dust \citep{mathis1977, draine1984}. The templates are parameterized by three parameters describing the shape of the SED over 1--$10^4~\mu$m: the shape ({\it duste\_gamma}; 0.1) and location ({\it duste\_umin; 1.0}) of the thermal dust emission bump in the IR SED, and the fraction of the total dust mass attributed to PAHs ({\it duste\_qpah}; 3.5). We included emission from circumstellar dust around thermally pulsating AGB stars \citep{villaume2015}.
Nebular emission based on {\sc cloudy} models from \citet{byler2017} was also assumed. 

The mass of the fitted stellar population was a free parameter, with a uniform prior between $10^{6}$ and $10^{10}$ $M_{\odot}$, but in the current study we were not interested in the stellar masses of the somewhat arbitrarily chosen regions, and will not discuss them in the rest of the paper.

While the observations are limited to wavelengths $\lesssim 3.6~\mu$m, such that there are no direct constraints on the dust emission, the SED-fitting results are marginalized over the unconstrained parameters such that the uncertainties are better accounted for than if a fixed dust attenuation model was assumed. The prior on the {\sc fsps} dust model parameter $\hat{\tau}_{\rm dust,~diffuse}$ used was uniform with a minimum of 0.0 and maximum of 2.0.

\begin{figure*}
\begin{center}
\includegraphics[scale=0.9]{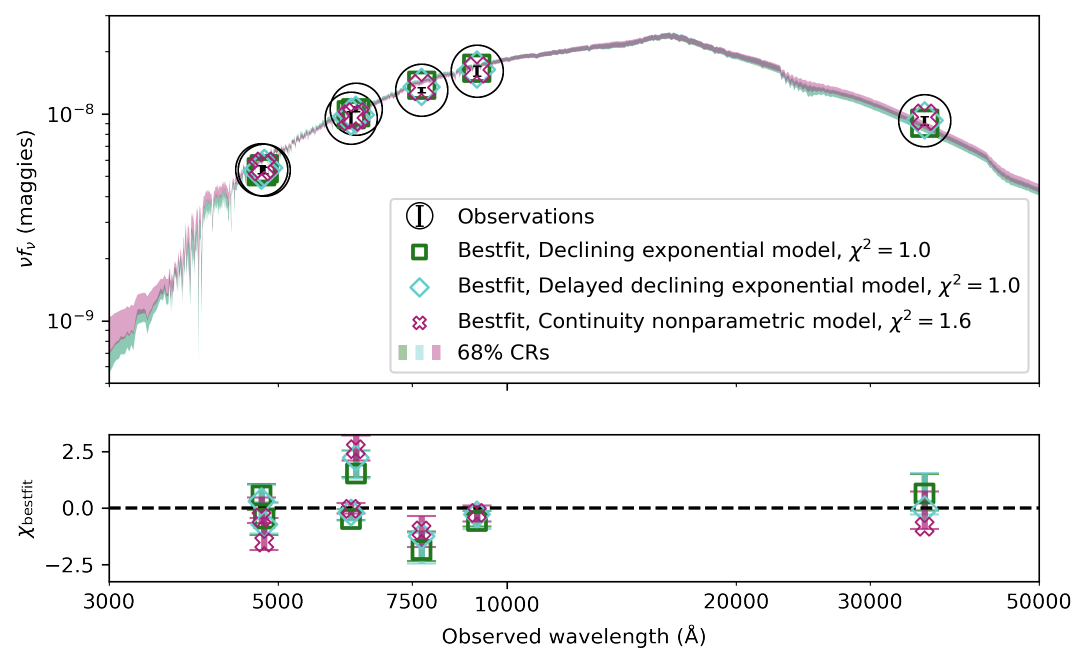}
\caption{Example of observations and fitting results for Region 1 with the three SFH priors. Markers indicate the photometry and the best-fit solutions. Shaded regions (errorbars) indicate the 68 per cent spread of the SEDs ($\chi^2$) drawn from the posteriors. A \citet{kriek2013} dust law was assumed.}
\label{fig2}
\end{center}
\end{figure*}

We have not seen any evidence for particularly blue colors or recent star formation in the ring-like tidal stellar stream around SPRC047, and therefore believe that exponentially declining SF models for SPRC047's stellar stream are plausible. Thus, three distinct SFH models, and three sets of SFH priors, were used:
\begin{enumerate}
    \item An exponentially declining $\tau$ model, $\mathrm{SFR} \approx \exp \left(-\frac{t - t_0}{\tau} \right)$, where $t > t_0$. This model is commonly adopted in the literature \citep[e.g.,][]{pandya2018}. The star formation rate increases instantaneously from zero to a maximum value at time $t_0$, after which it declines with an e-folding timescale $\tau$. This model best describes an SFH that is dominated by one epoch of star formation, but can otherwise lead to biased results \citep[\textit{e.g.},][]{carnall2019a}. The priors for the parameters follow $t_0~(\mathrm{Gyr}) \approx \mathrm{Uniform}(0.1, ~14.0)$, and $\tau~(\mathrm{Gyr}) \approx \mathrm{LogUniform}(0.1, ~10)$.\footnote{Using a prior $\tau / \mathrm{Gyr} \sim \mathrm{Uniform}(0.1, ~10)$ instead produces consistent results with those discussed in this work.}
    \item A delayed-$\tau$ model, $\mathrm{SFR} \approx (t - t_0)\exp \left(-\frac{t - t_0}{\tau} \right)$ that removes the discontinuity in the SFH at $t_0$ and allows for rising SFRs. The same priors for $t_0$ and $\tau$ as used in the $\tau$-model were assumed.
    \item A nonparametric model that assumes that the SFH can be described as a piece-wise constant function specifying the SFR in fixed time bins. Five time bins were used, the first bin from 1 year to 30~Myr, second from 30~Myr to 100 Myr, and the last bin from 0.85 times the age of the universe to the age of the universe. The rest of the time is split up between the remaining bins to have equal spacing in log($t$). This reflects the expected resolution of distinct stellar populations given the limited age constraints provided by the broadband observations. The nonparametric model is more flexible than the parametric $\tau$ models in that recent star formation does not preclude an underlying older stellar population (see \citealt{leja2019, lower2020}). That said, a prior is assumed for the relative SFR between adjacent time bins: a ``continuity'' prior that sets $\log \left( \mathrm{SFR}_n / \mathrm{SFR}_{n+1} \right) \approx \mathrm{StudentT}(\mu = 0, ~\sigma=0.3, ~\nu=2)$. This places a preference for constant SFHs, where the strength of the preference ($\sigma=0.3$) is tuned to match the dispersion in SFHs from simulated galaxies (see \citealt{leja2019} for details). This tuning was done for simulated galaxies, and has been used widely in the literature. However, there is no expectation that this tuning should definitely represent the object being studied in this work.
\end{enumerate}

The likelihood ($\mathcal{L}$) model used in {\sc prospector} follows a $\chi^2$ distribution such that
\begin{equation}
    \ln \mathcal{L} = - \frac{1}{2} \sum_{i=1}^{N} 
    \left[
    \ln \left( 2 \pi \sigma_i^2 \right) 
    + 
    \frac{ 
    \left(  ~
    f_i - f_i( \theta )
    ~\right)^2 
    }{
    \sigma_i^2
    }
    \right]
\end{equation}
where at a given photometric point each observed flux density, $f_{i}$, (with an associated uncertainty, $\sigma_i$) is compared to the model flux density, $f_i( \theta )$, given the model parameters $\theta$ \citep{johnson21}.

The SED models were sampled using a dynamic nested sampling approach (\texttt{dynesty}; \citealt{speagle20}).\footnote{Results were also obtained using ensemble Markov chain Monte Carlo (MCMC) sampling with \texttt{emcee} \citep{goodman10}, and they were consistent with the results produced by \texttt{dynesty}, and as presented in this work.} The stopping criterion for the sampling algorithm was chosen to ensure that the peak of the posterior distributions was well sampled. More specifically, we used a stopping criterion of $\Delta$ log $\hat{Z}$ = 0.05 \citep[see][]{speagle20} where $\hat{Z}$ is the evidence. The fitted free parameters and their uncertainties reported in this work reflect the median and its 68~per cent credible regions (CRs; they correspond to the 16$^\mathrm{th}$ to 84$^\mathrm{th}$ per cent range) of the posterior probability density functions, as the majority of the distributions are non-symmetric. The redshift $z$ of SPRC047 is known, and is fixed to 0.03117 in all the SED fits.

\section{Results} \label{sec:results}

\begin{figure*}
\begin{center}
\includegraphics[scale=0.7]{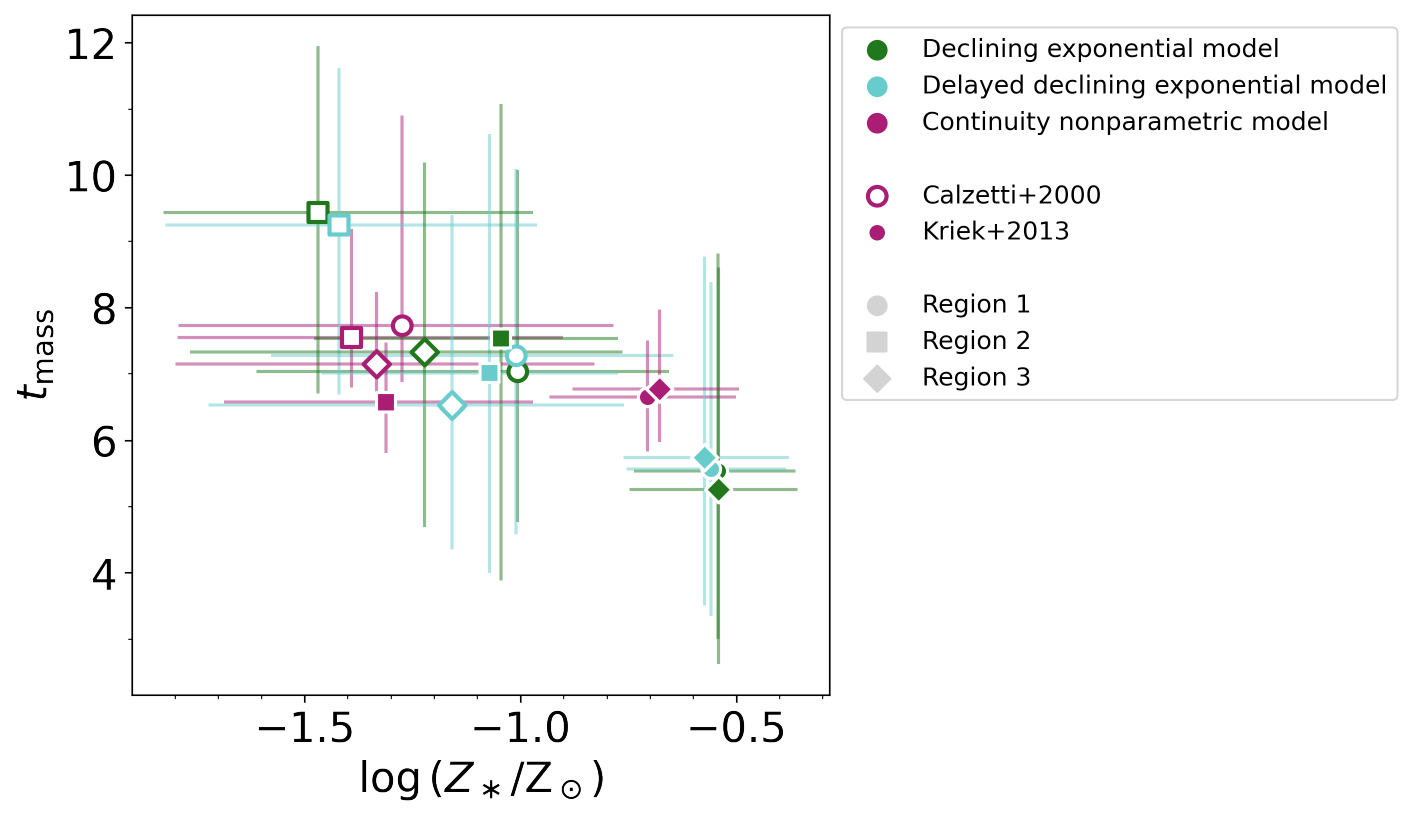}
\caption{Correlation of the mean weighted age and metallicity. All three regions (see Figure~\ref{figopt}) are shown with different symbols and colors.}
\label{age-metallicity-fig}
\end{center}
\end{figure*}

\begin{figure*}
\begin{center}
\includegraphics[scale=0.8]{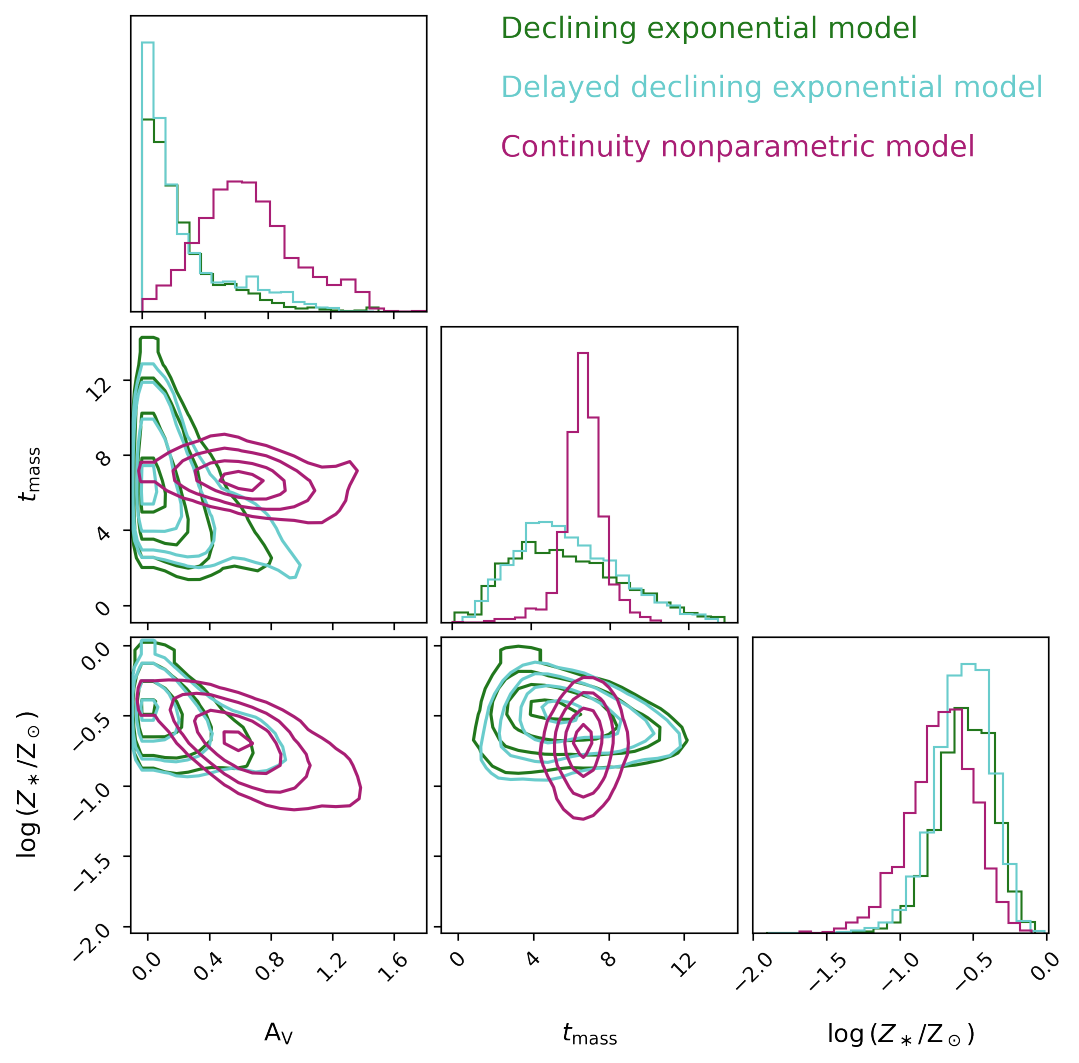}
\caption{Corner plot for Region 1. This plot compares the derived values for dust extinction $A_V$, $t_{\rm mass}$ (or mean mass-weighted age), and log($Z_{*}/Z_{\odot}$) and shows their correlations, The three studied different SFH models (labeled with different colors) were combined with the \citet{kriek2013} dust attenuation law. Contours are shown smoothed with an n = 1 Gaussian kernel.}
\label{cornerplot-fig}
\end{center}
\end{figure*}

\subsection{Morphology} \label{sec:morpho}

As shown in Figure~\ref{figopt}, the ring-like streams form an ellipse that has a projected major axis that is almost perfectly aligned in the north--south orientation. However, the brightest parts of the ellipse look off-centered with respect to the central edge-on galaxy, having a larger fraction of the ellipse on the northern side of the edge-on galaxy than on the southern side. In addition, there appears to be a faint branching of the ellipse on the northern side, where a ``loop'' closes the ellipse in such a way that approximately equally large parts of the ellipse are seen on both the northern and southern sides. These two northern stellar loops could either have been left behind by two separate disrupting progenitors or they could correspond to different stripping episodes of a single stream. Therefore, the morphology is that of a double-loop structure on the northern side of the edge-on galaxy, whereas the southern side only shows one half-loop. This probably means that the interaction that caused the formation of these loops is fairly recent,  and the structure is not really a typical polar-ring galaxy. However, \citet{reshetnikov15} included SPRC047 in their sample of polar ring galaxies and discussed the colors and the stellar masses of the rings, and the colors and SFRs of their host galaxies. Further evidence for a fairly recent merger is provided by \citet{combes13}, who detected molecular CO gas in their pointing towards SPRC047. Unfortunately, the gas was not spatially resolved between the ring and the central edge-on galaxy, and therefore, the association of the gas with the stellar stream is not certain.

The surface brightness of the loops is not smooth along the ring, but there are brighter areas near the major axis of the loops (near the northernmost and southernmost points of the bigger ellipse). There are also several point-like sources projected onto the ring-like stream, specifically on the western side. Two bright point-like sources on the loop are seen on the northwestern side of the larger loop, and one on the southwestern side of the loop. In addition, as seen in Figure~\ref{fig-deconv}, there is a point-like source just to the east of the northernmost point in the main loop in the IRAC 3.6 $\mu$m image. We tried fitting a galaxy-like SED to the first three point-like sources mentioned above, and were unable to obtain good fits. Therefore, we suspect that these point-like sources are all foreground stars. The nature of the fourth point-like object is not known, but we suspect that it is a foreground object as well. Therefore, there is no evidence of a remnant nucleus of a disrupted satellite galaxy that has formed the ring-like stellar stream.

The regions that we used for estimating stellar population parameters in the main loop excluded the point-like objects, so that the contribution of light from these faint foreground stars would be minimized. The fitted regions are discussed next.

\subsection{Photometry} \label{sec:photometry}

We measured integrated stellar emission in three regions of the ring-like tidal stellar stream around SPRC047, as shown in Figure~\ref{figopt}. The final measured integrated photometry in AB magnitudes, obtained with the {\sc SAOImageDS9} image viewer \citep{joye03}, is given in Table~\ref{table1}. We performed aperture photometry using the DS9 tool in irregularly shaped polygons that incorporated areas of the ring that did not have any bright foreground or background objects. The typical surface brightness at 3.6~$\mu$m in the measured areas of the ring-like tidal stellar stream is 25~AB mag~arcsec$^{-2}$. The typical stream colors in the measured apertures are $(g-r)_{\rm 0}$ $\approx$~0.6--0.7, $(g-i)_{\rm 0}$ $\approx$~0.7--1.0, $(g-z)_{\rm 0}$ $\approx$~1.1--1.2, and $(g-3.6)_{\rm 0}$ $\approx$~0.3--0.6. The $g-r$ colors are consistent with the findings of the stellar stream colors by \citet{miro2023}  and \citet{martinez2023b}, and redder than those of typical satellite dwarf galaxies.

To estimate how much PSF-aliased emission from the bright  central edge-on galaxy would have been added to the IRAC 3.6 $\mu$m photometry of the faint ring-like stellar stream at various distances from the central edge-on galaxy, we used a {\sc galfit} \citep[][]{Peng2002} model of the central edge-on galaxy in the $z$-band as an approximate proxy for its deconvolved structure in the IRAC 3.6 $\mu$m band, and then convolved this model with the IRAC 3.6 $\mu$m extended PSF from \citet{laine2016} to estimate the rough contribution of the aliasing of emission by the extended PSF in this band. The result was that the edge-on galaxy contributes about 1/3 of the light in those parts of the ring-like stream that are close to the plane of the edge-on galaxy and about 1/10 of the light in the ring-like stream at its farthest distance from the plane of the edge-on galaxy. This result shows how important a careful deconvolution, described above in Section \ref{sec:deconv}, is for obtaining realistic flux estimates in the various parts of the ring-like stream at 3.6 $\mu$m.

The PSF-aliased contribution from the central edge-on galaxy to the measured regions in the stream in the visual bands is much less significant. More quantitatively, we estimated the contribution from the bright central edge-on galaxy disk into our measured regions (Figure~\ref{figopt}) by modeling the central edge-on galaxy disk with ellipse fits. We then used the model to estimate the contribution of the PSF-aliased emission from the central edge-on galaxy disk to the regions where we measured the flux densities. The contribution from the central edge-on galaxy disk is less than the estimated uncertainties in the flux density in Table~\ref{table1} for Regions 2 and 3, and comparable to the uncertainties for Region 1. As expected, PSF-aliasing is most severe in the region closest to the edge-on galaxy.

\subsection{{\sc prospector} Models for the Tidal Stellar Stream}\label{sec:models}

The results from the models that we ran with {\sc prospector} are summarized in Table~\ref{table2}. We tabulate the 50th, and as uncertainty indicators, the 16th and 84th quantiles of the posterior probability distribution functions of the following free parameters: the mean mass-weighted age ($t_{\rm mass}$), $t_{\rm 0}$ that is the age since the first onset of star formation for the $\tau$ models, the metallicity (log[($Z/Z_{\odot}$)]), where $Z_{\odot}$ = 0.0142 \citep{asplund2009}, and the extinction in the visual band $A_{V}$, as derived from the dust model (see Section~\ref{sec:SED}). The mass of the fitted stellar population was a free parameter as well, as stated in Section~\ref{sec:SED}. We show results from our SED fits in Figures~\ref{fig2}--\ref{cornerplot-fig}.\footnote{We do not give statistical estimators of the goodness of the fits, such as $\chi^{2}$ or Bayes factors in the case of nested sampling because, e.g., the ``bestfit'' solution is just the set of parameter values for which the sampling algorithm happened to find the highest probability, and given that the evidence is tightly tied to the volume of the prior space that differs in all the prior assumptions that we have investigated. In addition to the differing prior assumptions (and prior volume), \citet{lawler2021} have shown that the Bayes factors may ``fail'' in the low S/N regime, such as in a faint stellar stream.} 

\begin{table*}
    \centering \footnotesize
    \caption{SED-fitting results from \texttt{dynesty} sampling. $t_{\rm mass}$ is the mass-weighted age. The tabulated values are the 50 (16, 84) per cent quantiles of the posterior. Some of the results from this Table are shown in Figure~\ref{age-metallicity-fig}.\label{table2}}
    \label{tab:results_main}
    \begin{tabular}{cc|cccc}
    \hline
        Reg. & SFH & \multicolumn{4}{c}{Dust model: \citet{kriek2013} } \\
        \hline
        ~ & ~ & $t_0$ (Gyr) & $t_\mathrm{mass}$ (Gyr) & $\log( Z_\ast / \mathrm{Z}_\odot )$ & $A_{V}$ (mag) \\
        \hline
        1 & $\tau$ & 6.76 (3.7, 10.7) & 5.6 (3.0, 8.9) & -0.54 (-0.75, -0.36) & 0.17 (0.05, 0.45) \\
        ~ & Delayed-$\tau$ & 7.3 (4.7, 10.8) & 5.6 (3.3, 8.4) & -0.56 (-0.75, -0.38) &  0.16 (0.04, 0.56) \\
        ~ & Continuity & \nodata & 6.6 (5.9, 7.5) & -0.70 (-0.93, -0.50) & 0.65 (0.38, 0.99) \\
        \hline
        ~ & ~ & \multicolumn{4}{c}{Dust model: \citet{calzetti2000} }  \\
        \hline
        ~ & ~ & $t_0$ (Gyr) & $ t_\mathrm{mass}$ (Gyr) & $\log( Z_\ast / \mathrm{Z}_\odot )$ & $A_{V}$ (mag) \\ 
        \hline
        ~ & $\tau$ & 8.0 (5.2, 11.6) & 7.0 (4.8, 10.1) & -1.00 (-1.61, -0.65) & 0.47 (0.20, 0.78) \\ 
        ~ & Delayed-$\tau$ & 8.7 (5.4, 12.3) & 7.3 (4.6, 10.1) & -1.01 (-1.58, -0.65) & 0.47 (0.19, 0.78)\\ 
        ~ & Continuity & \nodata & 7.7 (6.9, 10.9) & -1.28 (-1.79, -0.78) & 0.78 (0.45, 1.01) \\
        \hline
        ~ & ~ & \multicolumn{4}{c}{Dust model: \citet{kriek2013} }  \\
        \hline
        ~ & ~ & $t_0$ (Gyr) & $t_\mathrm{mass}$ (Gyr) & $\log( Z_\ast / \mathrm{Z}_\odot )$ & $A_{V}$ (mag) \\
        \hline
        2 & $\tau$ & 8.8 (5.1, 12.2) & 7.5 (3.8, 11.1) & -1.04 (-1.46, -0.78) & 0.27 (0.08, 0.67) \\
        ~ & Delayed-$\tau$ & 9.4 (5.8, 12.5) & 7.0 (4.1, 10.6) & -1.08 (-1.47, -0.78) & 0.31 (0.10, 0.71) \\
        ~ & Continuity & \nodata & 6.6 (5.8, 7.5) & -1.31 (-1.69, -0.97) & 0.82 (0.55, 1.10) \\ 
        \hline
        ~ & ~ & \multicolumn{4}{c}{Dust model: \citet{calzetti2000} } \\
        \hline
        ~ & ~ & $t_0$ (Gyr) & $t_\mathrm{mass}$ (Gyr) & $\log( Z_\ast / \mathrm{Z}_\odot )$ & $A_{V}$ (mag) \\
        \hline
        ~ & $\tau$ & 10.2 (7.3, 12.8) & 9.4 (6.7, 12.0) & -1.45 (-1.83, -0.97) &  0.45 (0.18, 0.62) \\ 
        ~ & Delayed-$\tau$ & 10.7 (7.7, 12.9) & 9.2 (6.7, 11.6) & -1.41 (-1.82, -0.96) & 0.44 (0.18, 0.63) \\ 
        ~ & Continuity & \nodata & 7.5 (6.8, 9.1) & -1.39 (-1.79, -0.92) & 0.62 (0.31, 0.82) \\
        \hline
        ~ & ~ & \multicolumn{4}{c}{Dust model: \citet{kriek2013} } \\
        \hline
        ~ & ~ & $t_0$ (Gyr) & $t_\mathrm{mass}$ (Gyr) & $\log( Z_\ast / \mathrm{Z}_\odot )$ & $A_{V}$ (mag) \\
        \hline
        3 & $\tau$ &  6.5 (3.3, 10.5)  & 5.3 (2.6, 8.6) & -0.54 (-0.74, -0.35) & 0.23 (0.07, 0.54) \\
        & Delayed-$\tau$ & 8.1 (4.6, 11.7)  & 5.8 (3.5, 8.8)  & -0.57 (-0.76, -0.37) & 0.21 (0.06, 0.60) \\
        & Continuity & \nodata & 6.8 (6.0, 8.0) & -0.68 (-0.88, -0.49) & 0.66 (0.42, 1.02)\\
        \hline
        ~ & ~ & \multicolumn{4}{c}{Dust model: \citet{calzetti2000} }  \\
        \hline
        ~ & ~ & $t_0$ (Gyr) & $t_\mathrm{mass}$ (Gyr) & $\log( Z_\ast / \mathrm{Z}_\odot )$ & $A_{V}$ (mag) \\
        \hline
        ~ & $\tau$ & 8.6 (5.3, 12.1) & 7.3 (4.7, 10.2) & -1.22 (-1.76, -0.77) & 0.64 (0.34, 0.90) \\
        ~ & Delayed-$\tau$ & 8.0 (5.2, 11.5)  & 6.5 (4.3, 9.4) & -1.16 (-1.72, -0.76) & 0.63 (0.33, 0.88) \\
        ~ & Continuity & \nodata & 7.2 (6.5, 8.2) & -1.33 (-1.80, -0.83) & 0.92 (0.59, 1.15) \\ 
    \end{tabular}
\end{table*}

\section{Discussion} \label{sec:discussion}

\subsection{Dependence of Results on Assumed Models}\label{sec:dependence}

The dust model and the SFH model (where the mass-weighted age is a latent parameter of the SFH model) were both allowed to vary in our SED fits (along with other free parameters), but they are degenerate. Figure~\ref{age-metallicity-fig} demonstrates that both the inferred metallicity and mass-weighted age are highly influenced by the assumed SFH and dust model. This implies that there is not sufficient information in the observations to constrain the SFH and the dust model parameters. Unfortunately, little is known about the nature and abundance of dust in stellar streams.

Our results indicate a distinct lack of recent star formation, suggesting that the ages of the regions are at least as old as 5 Gyr. With the S/N (~$\approx$~30) of the measured photometry, we do not expect to achieve a more accurate estimate of the ages of the stellar populations with any SFH model. Based on previous work, e.g., \citet{webb2022}, having both/either UV or mid-infrared observations in addition to spectroscopy (to break the age--dust--metallicity degeneracy) appears to be critical for constraining the dust content in low surface brightness targets. In addition, the ages from the parametric SFH models vary a lot. This suggests also that overall the results are noisier than the uncertainties in the measurements suggest. 

One consistent result is that Region 2 is the oldest (both in $t_{\rm 0}$ and in mass-weighted age). Surprisingly, the S/N value in this region is the lowest of the three regions. If we assume that the marginalized prior distribution for mass-weighted age describes the natural preference of the model, then the parametric models prefer younger ages than the posterior values. Where the S/N of the observations is lower, the posteriors are naturally more impacted by the prior distribution. The fact that the ages for Region 2 are in fact oldest is therefore curious.

We also tested for the weighting of the observations, as we have two $g$- and two $r$-bands in the input data (although, with slightly different band centers and widths). Thus, we reran the SED fits to the observations with all the SFH priors used in this work and using the \citet{kriek2013} dust models and only the CFHT $g$- and $r$-bands (leaving the Legacy Survey data out). The results for $t_{\rm mass}$, $t_{\rm 0}$ and metallicity were the same, within the uncertainties, as the values tabulated in Table~\ref{table2}, which used both the CFHT and Legacy Surveys data in the $g$- and $r$-bands. The fact that there is more data in the $g$- and $r$-bands makes us more confident in the correct determination of the flux densities at the corresponding wavelengths (the measured values are within the uncertainties of each other in each of the $g$- and $r$-bands). Therefore, we argue that attaching ``more weight'' to these bands and using all the available data is the best approach (corresponding to the results in Table~\ref{table2}). 

Fitting a nonparametric model with a prior that is less favorable to ``smooth'' SFHs (Dirichlet $\alpha$ $<$ 1 prior) would provide another set of fits to compare with, where we would expect them to produce ages that fall between the ages of the continuity and delayed-tau models shown in the current study. A Dirichlet prior with $\alpha$ $<$ 1 offers a prior that (in contrast to the continuity prior) does not care about ``smoothness'' at all. However, we would not expect the use of a Dirichlet prior to result in a quantitative improvement in the age determination of the stellar stream segments.

\begin{deluxetable*}{lllll}
\tablecaption{Input and output parameter values in the mock SED fit. \label{table3}}
\tablehead{
\colhead{Stellar mass} & \colhead{log($Z$/$Z_{\odot}$)} & \colhead{A$_{V}$} & \colhead{$t_{\rm 0}$ $\times$$10^{8}$} & \colhead{$\tau$} \\
\colhead{(M$_{\odot}$)} & \colhead{} & \colhead{(mag)} & \colhead{(yr)} & \colhead{(yr)}
}
\startdata
\multicolumn{5}{c}{Input values:} \\
1.0$\times$$10^{8}$ & $-1.1$ & 0.22 & 8.4 & 1.0 \\
\multicolumn{5}{c}{Fitted values:} \\
$9.27_{-0.66}^{+0.48}$$\times$$10^{7}$ & $-1.05_{-0.04}^{+0.04}$ & $0.22_{-0.00}^{+0.00}$ & $7.50_{-0.54}^{+0.55}$ & $0.88_{-0.06}^{+0.07}$ \\
\enddata
\end{deluxetable*}

\subsection{Tests with Mock Observations} \label{sec:mock tests}

A more accurate determination of the stellar population age and metallicity and other star formation parameters than what is possible with the current observations, as reported in this work, is highly desirable. This requires deeper observations with higher S/N ratios, and a larger spectral coverage than what is currently available for SPRC047. We can simulate the accuracy of stellar population parameter determination from such future high S/N and broad wavelength coverage observations with {\sc prospector} and {\sc fsps}.

In the near future, several observatories with large ground-based telescopes or telescopes in space, such as the Vera C. Rubin Observatory \citep[e.g.,][]{ivezic19}, Euclid \citep[e.g.,][]{laureijs2011}, Nancy Grace Roman Space Telescope \citep[e.g.,][]{akeson19} and the James Webb Space Telescope \citep[JWST; e.g.,][]{gardner2006} will be able to perform deep observations of low surface brightness targets in the sky. We made mock observations of a stellar stream portion similar to Region 1 that had broad-band data from FUV and near-UV (NUV; similar to GALEX FUV and NUV channels; \citeauthor[e.g.,][]{martin2005} \citeyear{martin2005}), visual to near-infrared $griz$ filters, similar to those of CFHT's MegaPrime instrument \citep{boulade2003}, Spitzer/IRAC 3.6, 4.5, and 5.8 $\mu$m channels, and mid-infrared 13, 15, 21, and 25 $\mu$m bands, similar to those of JWST's MIRI \citep[e.g.,][]{ressler2015} F1280W, F1500W, F2100W, and F2500W bands. We set the properties of the mock-observed stellar stream stellar population as given in Table~\ref{table3}. We used the exponentially declining SFH and the \citet{kriek2013} dust extinction law. Dust and nebular emission was also included. We set the S/N ratio to 100 (which is perhaps a little optimistic, and driven by the absolute instrument flux or photometric calibration accuracy), and added small offsets to the simulated broad-band photometry, drawn randomly from a normal distribution that used the noise value as the input $\sigma$, and ran {\sc prospector}. The free parameters were the metallicity, $t_{0}$, $\tau$, $A_{V}$\footnote{Not strictly true. See Equation~\ref{eqn:dust_diffuse_2}; the normalization of the diffuse dust attenuation curve was also a free parameter.}, and stellar mass. The median posterior values for these free parameters, together with their 16 per cent and 84 per cent uncertainties, are given in Table~\ref{table3}. 

We see that the dust content has been exactly determined using the S/N = 100 data and the exponentially declining SFH prior. However, even with pristine photometry we do not recover the correct SFH parameters. We also do not quite recover the stellar metallicity. The metallicity, $\tau$, and $t_{\rm 0}$ are all underestimated, which is not consistent with the expected age--metallicity degeneracy. It is known that younger stellar populations are intrinsically easier to recover, because the evolution of stellar SEDs is nonlinear. On the other hand, the ages of old stellar populations are mainly subject to the choice of an SFH model and prior \citep{leja2019,carnall2019a}. Note that the mock SFH was constructed in such a way that there was only one epoch of SF, and therefore the age effect is not investigated in this mock test. 

There is a degeneracy between $t_{\rm 0}$ and $\tau$. The log-uniform prior for $\tau$ prefers lower values, so it is not surprising that the recovered $\tau$ is biased low. It is harder to understand why $t_{0}$ is also biased low. Photometric data alone do not completely resolve the degeneracy between the SFH and metallicity, however, so we would expect their uncertainties to be much larger than those of $A_{V}$, as seen in the results (Table~\ref{table3}).

What we conclude from this experiment is that the SFHs of old galaxies are difficult to measure, even from pristine, high S/N mock photometric observations. And moreover, even the derived 16--84 per cent range of the posteriors of the fitted parameters may not include the true values of these parameters. One of the advantages of nonparametric models is that their uncertainties better reflect the parameter space of reasonable results, although a caveat of this point is that it is also subject to the choice of a commensurate SFH prior. That said, if the main interest is only in the dust content or metallicity, the SED fitting results are more robust.

In summary, the inferred SFH parameters for SPRC047 contain considerable uncertainty. The best constraint for the age of the stellar population in SPRC047 is the lack of any evidence of recent star formation. The derived age-related parameters are suspect, or at least very sensitive to any model assumptions.

\section{Conclusions \label{sec:conclusions}} 

We have run an extensive set of {\sc prospector} SED model fits to visual--NIR aperture photometry in three sections of the faint ring-like tidal stellar stream seen around the edge-on galaxy SPRC047. We took special care to eliminate the aliasing of the emission by the extended IRAC 3.6 $\mu$m PSF from the bright edge-on central galaxy onto the low surface brightness ring-like tidal stellar stream by performing a state-of-the-art deconvolution.

We have demonstrated that SED-fitting derived results for faint tidal stellar streams, using only visual--NIR broad-band photometry, are strongly model dependent. We have included a detailed description of all the technical aspects of the SED models in an effort to distinguish what information is constrained by the observations, or assumed, given the nature of the SED model.

Our conclusion is that using only visual--NIR photometry is insufficient if attempting to derive accurate ages, metallicities, and dust content estimates for low surface brightness objects such as the stellar stream of SPRC047. Future attempts to derive accurate parameters of the stellar population in such objects should include high S/N broad-band data from FUV to mid-infrared wavelengths. Better yet, spectroscopy would provide even more stringent constraints, but it is currently too expensive in terms of observing time to help in a study of a larger set of low surface brightness outer galaxy features, such as tidal stellar streams. Given how many streams have already been detected in nearby galaxies \citep[e.g.,][]{martinez2023b} and how many will be detected in upcoming large area deep surveys, such as the Rubin Observatory's Legacy Survey of Space and Time and the Euclid Wide Survey, our results motivate the need for spectroscopic observations with the upcoming 30-m class telescopes.

\section{Acknowledgments}

We are indebted to Ben Johnson for several discussions and suggestions regarding the {\sc prospector} SED fits. We are thankful to Rick Arendt for his help with the PSFs used for testing the magnitude of the extended PSF-aliased emission. We thank Vihang Mehta for a careful reading of the manuscript and for his many helpful suggestions. Thanks to David Shupe for his expert Python plotting help. This work was authored by an employee of Caltech/IPAC under Contract No. 80GSFC21R0032 with the National Aeronautics and Space Administration. DMD acknowledges financial support from the Talentia Senior Program (through the incentive ASE-136) from Secretar\'\i a General de  Universidades, Investigaci\'{o}n y Tecnolog\'\i a, de la Junta de Andaluc\'\i a. DMD acknowledges funding from the State Agency for Research of the Spanish MCIU through the ``Center of Excellence Severo Ochoa'' award to the Instituto de Astrof{\'i}sica de Andaluc{\'i}a (SEV-2017-0709) and project (PDI2020-114581GB-C21/AEI/10.13039/501100011033). KW was supported by the National Sciences and Engineering Research Council of Canada (NSERC) PGS award. R.B.G was funded by the Universidad Internacional de la Rioja (UNIR) Research Project ``ADELA: Aplicaciones de Deep Learning para Astrof\'{i}sica,'' no. B0036-2223-1199 624-ADELA, and by the Call for Grants for Research Stays Abroad 2022/2023 of UNIR. S.P. acknowledges support from the Mid-Career Researcher Program (No. RS-2023-00208957) through the NRF of Korea. M.S acknowledges funding from the German Science Foundation DFG, within the Collaborative Research Center SFB1491 ``Cosmic Interacting Matters -- From Source to Signal.'' This work was partly done using GNU Astronomy Utilities (Gnuastro, ascl.net/1801.009) version 0.19. Work on Gnuastro has been funded by the Japanese Ministry of Education, Culture, Sports, Science, and Technology (MEXT) scholarship and its Grant-in-Aid for Scientific Research (21244012, 24253003), the European Research Council (ERC) advanced grant 339659-MUSICOS, the Spanish Ministry of Economy and Competitiveness (MINECO, grant number AYA2016-76219-P) and the NextGenerationEU grant through the Recovery and Resilience Facility project ICTS-MRR-2021-03-CEFCA. This research made use of Photutils, an Astropy package for detection and photometry of astronomical sources \citep{larry_bradley_2023_7946442}. This work is based in part on observations made with the Spitzer Space Telescope, which was operated by the Jet Propulsion Laboratory, California Institute of Technology under a contract with NASA. Based on observations obtained with MegaPrime/MegaCam, a joint project of CFHT and CEA/DAPNIA, at the Canada-France-Hawaii Telescope (CFHT) which is operated by the National Research Council (NRC) of Canada, the Institut National des Science de l'Univers of the Centre National de la Recherche Scientifique (CNRS) of France, and the University of Hawaii. The observations at the Canada--France--Hawaii Telescope were performed with care and respect from the summit of Maunakea which is a significant cultural and historic site. This research has made use of the NASA/IPAC Extragalactic Database (NED), which is operated by the Jet Propulsion Laboratory, California Institute of Technology, under contract with the National Aeronautics and Space Administration. This research has made use of NASA's Astrophysics Data System Bibliographic Services.

The Legacy Surveys consist of three individual and complementary projects: the Dark Energy Camera Legacy Survey (DECaLS; Proposal ID \#2014B-0404; PIs: David Schlegel and Arjun Dey), the Beijing-Arizona Sky Survey (BASS; NOAO Prop. ID \#2015A-0801; PIs: Zhou Xu and Xiaohui Fan), and the Mayall z-band Legacy Survey (MzLS; Prop. ID \#2016A-0453; PI: Arjun Dey). DECaLS, BASS and MzLS together include data obtained, respectively, at the Blanco telescope, Cerro Tololo Inter-American Observatory, NSF's NOIRLab; the Bok telescope, Steward Observatory, University of Arizona; and the Mayall telescope, Kitt Peak National Observatory, NOIRLab. Pipeline processing and analyses of the data were supported by NOIRLab and the Lawrence Berkeley National Laboratory (LBNL). The Legacy Surveys project is honored to be permitted to conduct astronomical research on Iolkam Du'ag (Kitt Peak), a mountain with particular significance to the Tohono O'odham Nation. 

NOIRLab is operated by the Association of Universities for Research in Astronomy (AURA) under a cooperative agreement with the National Science Foundation. 

This project used data obtained with the Dark Energy Camera (DECam), which was constructed by the Dark Energy Survey (DES) collaboration. Funding for the DES Projects has been provided by the U.S. Department of Energy, the U.S. National Science Foundation, the Ministry of Science and Education of Spain, the Science and Technology Facilities Council of the United Kingdom, the Higher Education Funding Council for England, the National Center for Supercomputing Applications at the University of Illinois at Urbana-Champaign, the Kavli Institute of Cosmological Physics at the University of Chicago, Center for Cosmology and Astro-Particle Physics at the Ohio State University, the Mitchell Institute for Fundamental Physics and Astronomy at Texas A\&M University, Financiadora de Estudos e Projetos, Fundacao Carlos Chagas Filho de Amparo, Financiadora de Estudos e Projetos, Fundacao Carlos Chagas Filho de Amparo a Pesquisa do Estado do Rio de Janeiro, Conselho Nacional de Desenvolvimento Cientifico e Tecnologico and the Ministerio da Ciencia, Tecnologia e Inovacao, the Deutsche Forschungsgemeinschaft, and the Collaborating Institutions in the Dark Energy Survey. The Collaborating Institutions are Argonne National Laboratory, the University of California at Santa Cruz, the University of Cambridge, Centro de Investigaciones Energeticas, Medioambientales y Tecnologicas-Madrid, the University of Chicago, University College London, the DES--Brazil Consortium, the University of Edinburgh, the Eidgenossische Technische Hochschule (ETH) Zurich, Fermi National Accelerator Laboratory, the University of Illinois at Urbana--Champaign, the Institut de Ciencies de l’Espai (IEEC/CSIC), the Institut de Fisica d’Altes Energies, Lawrence Berkeley National Laboratory, the Ludwig Maximilians Universitat Munchen and the associated Excellence Cluster Universe, the University of Michigan, NSF’s NOIRLab, the University of Nottingham, the Ohio State University, the University of Pennsylvania, the University of Portsmouth, SLAC National Accelerator Laboratory, Stanford University, the University of Sussex, and Texas A\&M University.

BASS is a key project of the Telescope Access Program (TAP), which has been funded by the National Astronomical Observatories of China, the Chinese Academy of Sciences (the Strategic Priority Research Program “The Emergence of Cosmological Structures” Grant \# XDB09000000), and the Special Fund for Astronomy from the Ministry of Finance. The BASS is also supported by the External Cooperation Program of Chinese Academy of Sciences (Grant \# 114A11KYSB20160057), and Chinese National Natural Science Foundation (Grant \# 12120101003, \# 11433005).

The Legacy Survey team makes use of data products from the Near-Earth Object Wide-field Infrared Survey Explorer (NEOWISE), which is a project of the Jet Propulsion Laboratory/California Institute of Technology. NEOWISE is funded by the National Aeronautics and Space Administration.

The Legacy Surveys imaging of the DESI footprint is supported by the Director, Office of Science, Office of High Energy Physics of the U.S. Department of Energy under Contract No. DE-AC02-05CH1123, by the National Energy Research Scientific Computing Center, a DOE Office of Science User Facility under the same contract; and by the U.S. National Science Foundation, Division of Astronomical Sciences under Contract No. AST-0950945 to NOAO.

\vspace{5mm}

\facilities{Spitzer, CFHT, Blanco}

\software{astropy \citep{astropy:2013,astropy:2018,astropy:2022}; numpy \citep{harris20}, scipy \citep{virtanen20}, emcee \citep{goodman10}, dynesty \citep{speagle20}, sedpy \citep{johnson19}, HDF5 \citep{hdf5}, h5py \citep{collette14}, {\sc fsps} \citep{conroy09,conroy10}, {\sc prospector} \citep{johnson21}, python-{\sc fsps} \citep{foreman14}, SAOImageDS9 \citep{joye03}, NoiseChisel \citep{gnuastro,noisechisel19}, IDLAWMLE \citep{baena:2013}, Photutils \citep{larry_bradley_2023_7946442}}

\bibliography{SPRC047_paper_resubm}{}
\bibliographystyle{aasjournal}

\end{document}